# Hydrogen Storage on Transition-Metal-Decorated Nitrogen-Modified Carbon Nanoribbons

Gom Dorji[1], Amrutha M[2], Brahmananda Chakraborty[3,4*],Sonam Peden[1], Syed Faraz Hasan[1,5], Shabbir Ahmad[6],and Tanveer Hussain[1*]

[1]School of Science and Technology, University of New England, Armidale, New South Wales, 2351, Australia

[2]Department of Physics, Karpagam Academy of Higher Education (Deemed to be University), Coimbatore-641021, India

[3]High Pressure & Synchrotron Radiation Physics Division, Bhabha Atomic Research Centre, Trombay, Mumbai 400085, India.

[4]Homi Bhabha National Institute, Anushaktinagar, Mumbai 400094, India

[5]Directorate of Research Services, University of Canberra, Bruce ACT 2617, Australia

[6]Queensland Alliance for Agriculture and Food Innovation, The University of Queensland, Brisbane, QLD, Australia

[*]Corresponding authors: brahma@barc.gov.in, tanveer.hussain@une.edu.au

## Abstract

The recently synthesised carbon nanoribbons (CNRs) were investigated for hydrogen ($H_2$) storage using first-principles density functional theory calculations. Pristine CNRs exhibited weak $H_2$ adsorption, therefore, the host structure was modified by substituting C–H edges with 12 N atoms, followed by Mn and Y doping to enhance $H_2$ binding. The maximum of 5 metal atoms were able to be accommodated on 12N+CNRs materials. Electronic structure analysis revealed strong orbital hybridization, while binding energy confirmed the structural stability and the Bader charge analysis shows quantity of charge transferred. The average $H_2$ adsorption energies were calculated to be −0.40 eV/$H_2$ for the Mn-doped system and −0.25 eV/$H_2$ for the Y-doped system, which are within the desirable range for reversible hydrogen storage. The maximum theoretical gravimetric storage capacities at 0 K reached 7.48 wt% for the Mn-doped system and 6.55 wt% for the Y-doped system. Under practical operating conditions (30 atm and 298.15 K), the storage capacity of the Y-doped system decreased to 6.04 wt%, whereas the Mn-doped system maintained its full storage capacity of 7.48 wt%.

Thermodynamic analysis further indicated that $H_2$ adsorption is favoured at low temperatures and high pressures, while desorption becomes feasible at elevated temperatures and lower pressures. These findings demonstrate that both transition-metal-doped CNRs satisfy the key U.S. Department of Energy (DOE) requirements for reversible $H_2$ storage and are promising candidates for practical $H_2$storage applications. The ability to exhibit such high levels of reversible $H_2$ storage under near-ambient conditions is another indication of their potential to reduce the energy costs of $H_2$ storage and thus improve the economic feasibility of the $H_2$ supply chain.

## 1. Introduction.

Worldwide energy demand is increasing by about 2% each year, with fossil fuels still accounting for the majority of energy production and contributing significantly to global greenhouse gas emissions [1]. Clean renewable energy sources have intermittent outputs, making it challenging to match energy supply with demand when needed [2-6]. $H_2$ is reported to have a higher energy density of 143 MJ $kg^{-1}$ compared to other fossil fuels like methane, natural gas, and kerosene [7]. However, at present, $H_2$ storage technology lags behind production and application due to its small size, low density, and gaseous nature. The current $H_2$ storage technologies include high-pressure gaseous $H_2$ storage [8], liquid $H_2$ storage [9] and solid $H_2$ storage [10]. The pressurised tank storage system faces challenges such as system weight, high cost, and low volumetric density (40 kg/$m^3$ at 70 MPa) [11]. The storage of liquid $H_2$ requires cryogenic temperature (-253ºC), resulting in consuming its own stored energy of ~30 to 40%, and also the boil-off loss of ~1% per day [12].These conventional methods are difficult to implement at large scale due to their high energy consumption, infrastructure requirements, and safety concerns, increasing the overall cost of $H_2$. Therefore, developing efficient solid-state $H_2$ storage materials is essential for reducing storage costs and improving the economic viability and large-scale deployment of $H_2$-based energy systems.

The other alternative is storing $H_2$ in solid materials. Solid-state $H_2$ storage (SSHS) methods include metal hydride $H_2$ storage, metal organic framework (MOF) materials, and carbonaceous materials [13]. SSHS, due to high safety, high storage density, higher energy efficiency, and reversibility, finds application in the prospects of transportation and energy industries and can promote large-scale applications [14].

MOF featuring large surface area, porosity, and tunable structure has been widely used in the field of gas adsorption, separation, and in the field of $H_2$ storage. However, in $H_2$ storage applications, $H_2$ uptake in MOFs depends on weak van der Waals interactions, these materials suffer from poor gas retention at moderate temperatures [15]. To improve their storage performance, various MOFs have been functionalised with alkali metals. At 77 K and 0.1 MPa, the $H_2$ storage capacity of MOF-5 increased from 2.49 wt% to 3.09, 4.18, and 4.23 wt% with Li, Na, and K functionalisation, respectively [16]. Similarly, Li-$Cu_3(BTC)_2$ and Li-MIL-101(Cr) achieved 3.50 and 3.39 wt%, compared with 2.41 and 2.37 wt% for their pristine counterparts [17]. However, under more practical conditions, the storage capacity remains low; IRMOF-9 achieved only 0.35 wt% at 298 K and 10 MPa, while MOF-C30 achieved 1 wt% at 300 K and 10 MPa, increasing to 5 wt% after Li doping [18, 19]. Overall, despite improved $H_2$ uptake through metal functionalisation, MOFs still suffer from poor $H_2$ retention and low storage capacity under practical conditions, limiting their ability to meet US DOE targets.

Another storage material is metal hydrides (MHs), which are highly compact and efficient volumetric $H_2$ storage through chemisorption, where $H_2$ molecules dissociate and form stable metal–$H_2$ (M–H) bonds. The paper [20] reported that $LiAlH_4$ can store $H_2$ of 10.54 wt% at 160$^{o}$C temperature. Similarly, the other metal hydrides like $NaAlH_4$, $KalH_4$, $Mg(AlH_4)_2$ and $Ca(AlH_4)_2$ can give storage capacities of 7.41, 5.71, 9.27 and 7.84 wt% at 210, 300, 110 and 130 $^{o}$C temperature, respectively [21]. Although MH systems can achieve comparatively high gravimetric storage densities relative to framework materials, their large-scale application is limited by slow $H_2$ desorption kinetics and inadequate structural reversibility. The other materials considered for $H_2$ storage studies are carbon nanostructures, as discussed in this study.

Recent studies have identified carbon nitride as a promising two-dimensional (2D) material for $H_2$ storage due to its low density, high porosity, large specific surface area, tunable structure, and excellent chemical stability [22-24]. For instance, Li-decorated CN and $C_2N$ monolayers were reported to achieve $H_2$ storage capacities of 7.50 and 7.27 wt%, respectively, with average adsorption energies of −0.17 and −0.13 eV/$H_2$, indicating favourable adsorption characteristics [25]. Similarly, a Li-functionalized $C_3N_6$ monolayer was found to exhibit a $H_2$ gravimetric capacity of 7.84 wt% with an adsorption energy of −0.22 eV/$H_2$ [26]. In our previous study, we investigated Mg-, K-, and Ca-decorated $C_3N_2$ monolayers using the DFT-D3 approach. The results showed that the $C_3N_2$ substrate could stably accommodate up to four metal dopants, each exhibiting a binding energy higher than its corresponding cohesive energy, thereby preventing metal clustering. Under fuel-cell operating conditions, the Mg-, K-, and Ca-functionalized $C_3N_2$ systems achieved gravimetric $H_2$ storage capacities of 9.47, 5.96, and 6.57 wt%, respectively [23]. Hussain et al. investigated the $H_2$ storage performance of boron-graphdiyne (BGDY) functionalized with light metal atoms using density functional theory (DFT) calculations. Their results indicated that the incorporation of light metal dopants significantly enhanced the $H_2$ adsorption capability of BGDY, enabling the system to achieve $H_2$ storage capacities that satisfy the targets established by the U.S. DOE [27]. Kadhim et al. studied transition-metal (TM)-functionalised ψ-graphene for $H_2$ storage using DFT. The TM atoms (Sc, Ti, and V) bonded strongly with ψ-graphene, with binding energies of 4.38, 4.11, and 2.44 eV for Sc, Ti, and V, respectively. TM-decorated ψ-graphene adsorbed six $H_2$ molecules, with adsorption energies of −0.31, −0.37, and −0.41 eV per $H_2$ molecule for Sc, Ti, and V, respectively [28]. In another study [29], TM-doped D-doped single-vacancy graphene was investigated for $H_2$ storage. The simulation results showed that Sc, Ti, and V adsorbed 5, 5, and 3 $H_2$ molecules, respectively, with adsorption energies ranging from −0.428 to −0.620 eV per $H_2$ molecule.

Motivated by the potential of carbon-based nanostructures, carbon nanoribbons (CNRs) are also synthesised and studied for potential $H_2$ storage. The nitrogen-modified CNRs with a 4-5-6-8-membered ring material of a one-dimensional material were considered for $H_2$ storage [30]. This CNRs structure is enhanced for $H_2$ molecule adsorption by doping transition metals such as scandium (Sc), titanium (Ti), and vanadium (V), and obtained the $H_2$ storage capacity of 6.10, 5.49 and 5.25 wt%, respectively, with favourable adsorption energies. In this work, CNRs of 4-5-6-8 rings of 52 C atoms, passivated with 16 $H_2$ atoms, are considered for $H_2$ storage. To further enhance the $H_2$ adsorption, two transition metals (TM) are selected, namely Mn and Y. Using DFT, the structure stability, charge transfer, binding energies, and related calculations are presented in a later section.

## 2. Computational methodology.

The density functional theory (DFT) calculations were carried out using the Vienna Ab initio Simulation Package (VASP) [31, 32]. The generalized gradient approximation (GGA) with the Perdew-Burke-Ernzhorf (PBE) functional was used to describe the electron exchange-correlation interaction [33]. The plane-wave cutoff energy (ENCUT) was set to 500 eV, according to the maximum recommended value of the PAW-PBE potentials used in the POTCAR files. The electronic minimization was converged with an energy difference criterion of $10^{-6}$eV between successive iterations, while the structural relaxation was performed until the residual forces were below 0.01 eV/Å with a maximum of 1500 ionic steps allowed to ensure complete convergence. The van der Waals interactions in the system were accounted for using the DFT-D3 method [34, 35]. Since the transition metal (TM)-doped CNRs are one-dimensional systems, the Brillouin zone was sampled using a Monkhorst–Pack k-point grid of 11x1x1 for structural optimization, and a denser k-point grid was adopted for accurate electronic structure calculations. The charge transfer characteristics between the TM dopants and CNRs were analysed using Bader charge analysis performed on the converged static charge density [36, 37].

Binding energies ($E_b$) between CNRs and TM are calculated using Eq. (1). The $H_2$ adsorption ($E_{ads}$) on TM-doped CNRs is calculated using Eq. ( 2), while the charge density difference in the system is studied using Eq. (3) [23].

$$E_b = \frac{E_{(12N+CNRs+mTM)} - E_{12N+CNRs} - mE_{(TM)}}{m} \quad (1)$$

$$E_{ads} = \frac{E_{(12N+CNRs+5TM+nH_2)} - E_{(12N+CNRs+5TM)} - nE_{H_2}}{n} \quad (2)$$

$$\Delta\rho = \rho_{(12N+CNRs+5TM)} - \rho_{(12N+CNRs)} - \rho_{(5TM)} \quad (3)$$

In Eq. (1), the $E_{(12N+CNRs+5TM)}$ represents the total system energy of TM-doped CNRs, $E_{(12N+CNRs)}$ represents the pure sheet energy of CNRs, $E_{(TM)}$ represents the total energy of TM, and m is the number of TM doped in CNRs. In Eq. ( 2), $E_{(12N+CNRs+5TM+nH_2)}$ is the total energy of hydrogenated TM-doped CNRs, and $nE_{H_2}$ is the isolated energy of n $H_2$ molecules. In Eq. (3) the right-hand side's first term represents the total charge of 5TM -doped CNRs, the second term represents the total charge of bare CNRs, and the third term gives the isolated charge of 5TM. The storage capacity is calculated using Eq. (4), where $nH_2$ is the total molecular mass of $H_2$, and 12N+CNRs + 5TM is the total molecular mass of the transition metal-decorated material.

$$wt\% = \frac{nH_2}{nH_2 + 12N + CNRs + 5TM} x\,100 \quad (4)$$

To examine the practical applicability of TM-doped 12N+CNRs, the desorption temperature is calculated using the Van't Hoff equation given by Eq. (5).

$$T_d = |E_{ads}| \Big/ \left[K_B\left(\frac{\Delta S}{R} - \ln P\right)\right] \quad (5)$$

Boltzmann constant is represented by the term $K_B$, R is the universal gas constant, and ΔS is the entropy change from gas to liquid for $H_2$ (75.44 J $mol^{-1}K^{-1}$), adopted from the reported value for $H_2$ adsorption on surfaces, reflecting partial loss of translational entropy upon adsorption [38-40] and P represents the equilibrium pressure of 1 atmosphere. The thermodynamic analysis of $H_2$ uptake mechanism is represented by Eq. (6).

$$N_{eff}(T,P) = \left[\frac{Z-1}{Z}\right] * N_o \quad (6)$$

Where Z is the grand conical partition function given by Eq. (7). $N_o$ represents the number of adsorbed $H_2$ molecules at 0.0 K obtained via VASP simulation.

$$Z(E_{ads},T,P) = 1 + \sum_{i=1}^{n} e^{-\left(\frac{E^i_{ads} - \mu_{H_2}}{K_B T}\right)} (7)$$

Where $E^i_{ads}$ is the average adsorption energy of the $i^{th}$ $H_2$ molecule, and 'i' index starts from 5, 10, 20, 30, and 40. Variable 'n' defines upper limits of $H_2$ molecules stored in decorated 12N+CNRs with $E_{ads}$ greater than -0.15 eV/$H_2$. $\mu_{H_2}$ is the gas phase chemical potential calculated using Eq. (8).

$$\mu_{H_2}(P,T) = \Delta H(T,P) - T\Delta S(T,P) + K_B T \ln\frac{P}{P_o} \quad (8)$$

Where $\Delta H - T\Delta S$ is Gibbs free energy calculated using the Shomate equation from [41]. ΔH is the standard enthalpy change of $H_2$ in kJ $mol^{-1}$, calculated using Eq. S1 (supplementary equation 1), ΔS is the standard entropy change calculated using Eq. S2 (supplementary equation 2) in $Jmol^{-1}K^{-1}$, both referenced to 298.15 K. Here, P is the operating $H_2$ pressure, while $P_o$ is the standard reference pressure (1 bar). The ratio $P/P_o$ represents the relative $H_2$ pressure and accounts for the pressure dependence of the chemical potential.

## 3. Results and discussion

### 3.1 Structural and electronic properties of CNRs

The one-dimensional CNR consists of 52 carbon atoms and 16 H atoms. The 16 H atoms are used to passivate the dangling bonds of carbon when the unit cell is formed. These $H_2$ atoms are not accounted for in any $H_2$ storage calculations. These CNRs have been experimentally synthesised by Faming Kang et al. in 2023 [42]. This CNRs consists of 4, 5, 6 and 8 carbon-membered rings with 2, 4, 10, and 1 number, respectively. CNRs are premier candidates for $H_2$ storage because their ultra-high specific surface area maximizes molecular contact, while their lightweight carbon framework minimizes dead weight to boost gravimetric capacity [43]. Unlike closed nanotubes, CNRs feature highly reactive, open edges that easily host nitrogen functionalisation and metal dopants without disrupting the core lattice. The C-C bond lengths vary from 1.38 to 1.48 Å, which agrees with previously reported data [30, 43]. The spin-polarized projected electronic density of states (PDOS) for pristine CNRs is presented in Figure 2 (a). The pristine CNRs are planar and semiconducting with a band gap of 0.5 eV, which agrees well with the literature, which has mentioned a band gap ranging from 0 to 0.5 eV [43]

The pristine CNRs are simulated for $H_2$ adsorption at different locations mentioned as A, B, C, D, E, F, and G in Figure 1 (a). The $H_2$ adsorption energies are calculated using Eq. ( 2) and presented in Table 1S (supplementary information). One-site $H_2$ adsorption is also presented in Figure 1S (supplementary information). It is observed that pristine CNRs have -0.16 eV/$H_2$ adsorption energies for all selected sites, which does not satisfy the adsorption energy window prescribed by the US DOE standard. To enhance $H_2$ adsorption, TMs are doped on CNRs. The results show that the most stable sites' binding energies for CNRs+Mn and CNRs+Y are -2.01 and -3.73 eV, respectively, much smaller than the cohesive energy, indicating the possibility of metal cluster formation. To prevent this, 12 N atoms are replaced with C-H edges to alter the local charge distributions and to study$H_2$adsorption. The bond length of C-N and C-C ranges from 1.33 -1.38 Å and 1.38-1.66 Å, respectively, which is in line with the literature [44]. The PDOS in Figure 2(b) shows strong hybridization between C and N atoms with minor orbital contribution from H atoms. There still exists a band gap of 0.5 eV and the system exhibits semiconducting. Figure 1 (b) shows different sites for $H_2$ adsorption, and simulated adsorption energies are presented in Table 2S (supplementary information). In all cases, the 12 N-atom-added system performs better than the pristine

CNRs; however, the $H_2$ storage capacity is still less than the required US DOE standard [45]. Thus, transition metal doping is considered for further analysis.

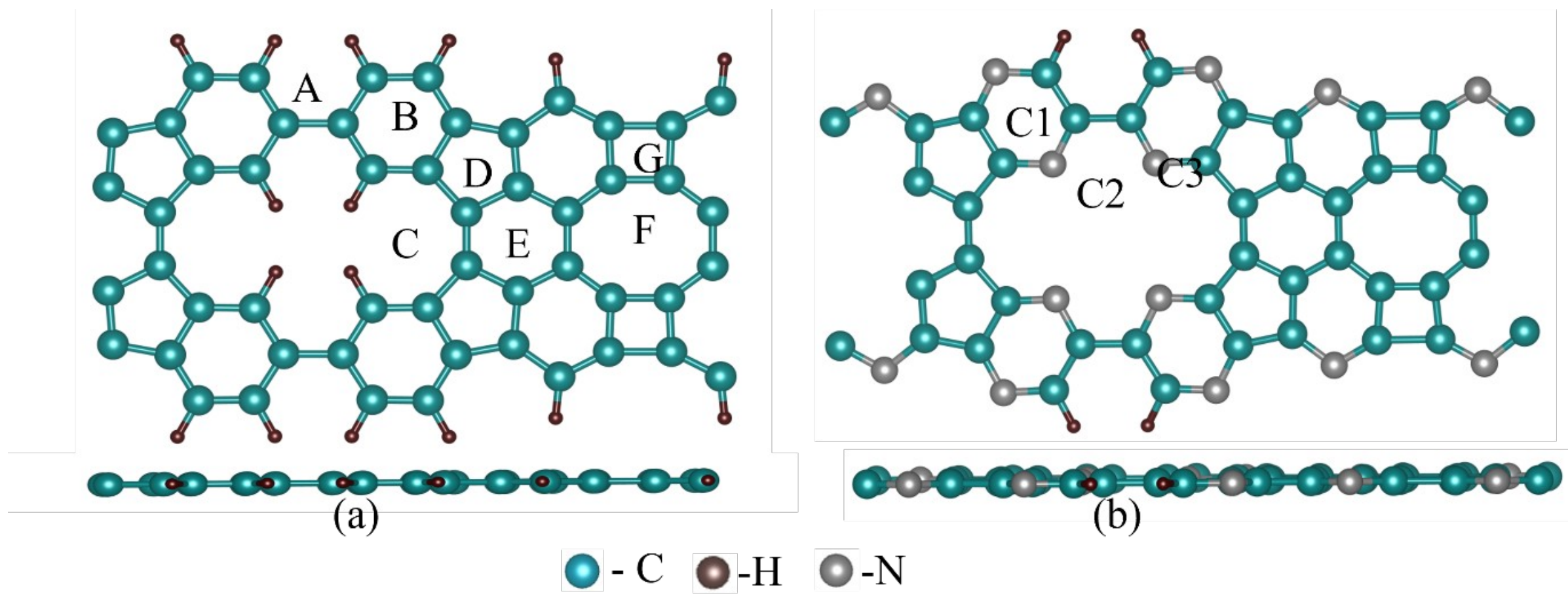

Figure 1. (a) Pristine CNRs and various sites for $H_2$ adsorption. (b) Nitrogen-substituted CNRs and different sites for $H_2$ adsorption.

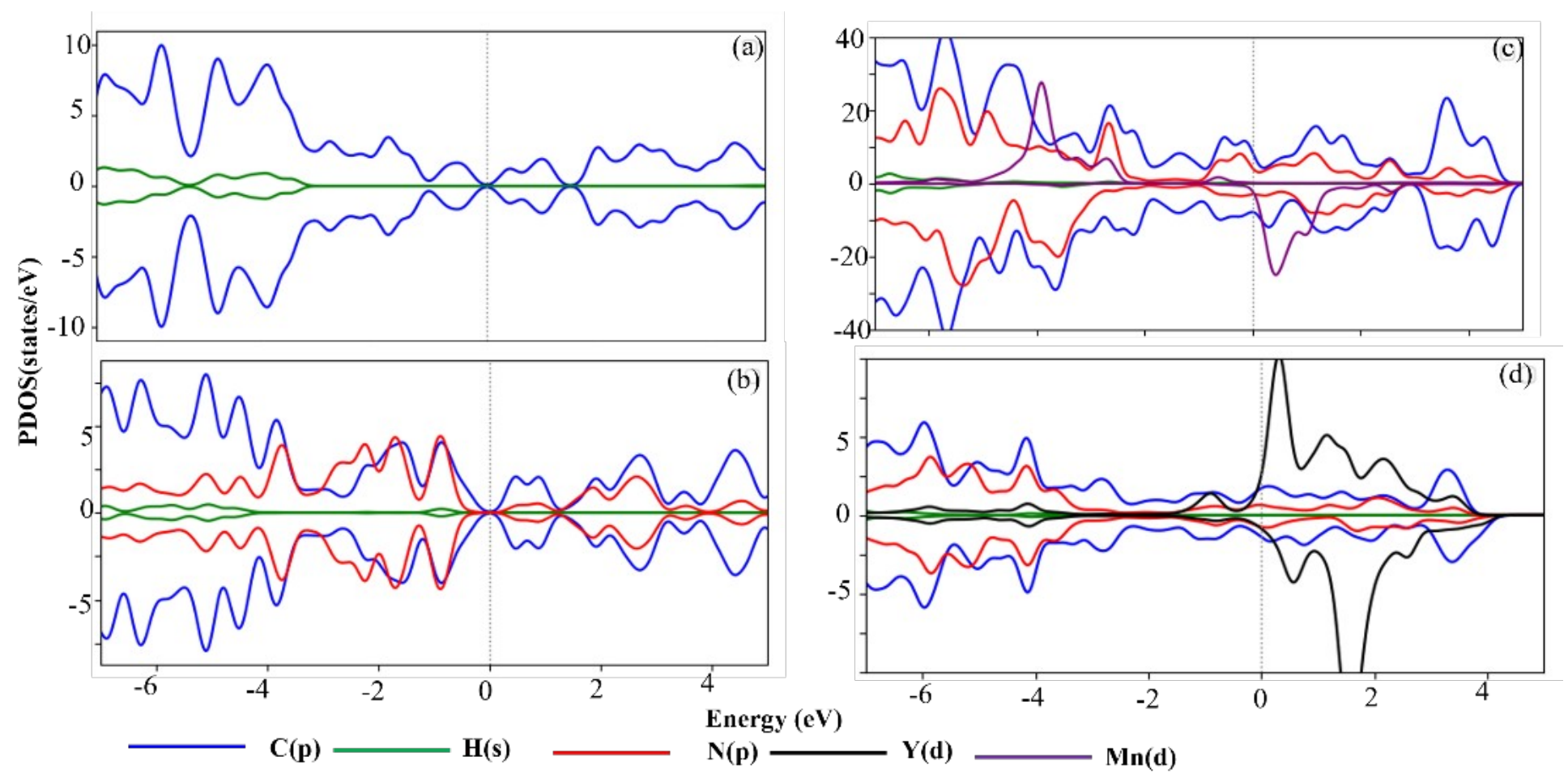

Figure 2. Projected density of states for (a) pure CNRs, (b) 12N+CNRs, (c) 12N+CNRs+5Mn, (d) 12N+CNRs+5Y. In (c), the PDOS of C, N, and H were scaled up by 7 times to make plot visible with respect to Mn PDOS.

### 3.2 Transition metal doping in 12N+CNRs

The 12N-replaced C-H CNRs are now optimized for transition-metal (TM) doping. TMs are selected as a functionalisation agent due to their partially filled d-orbitals, which helps in Kubas interaction with $H_2$ molecules. This will optimise the $H_2$ adsorption between the required energy ranges (-0.2 to -0.6 eV) necessary for reversible ambient condition storage [46]. The best position of TM dopants is selected based on the strongest binding energy.

Table 1 shows the binding energy of the maximum doping concentration compared with the cohesive energies to rule out TM cluster formation. It is observed that the binding energy of 12N+CNRs+5Mn is slightly weaker than the cohesive energy of bulk Mn. To further confirm this, the distance between Mn-Mn of simulation results is also compared with the bulk solid-state Mn-Mn distance, and it is observed that the literature distance reported is 2.25 Å [47], much shorter than the simulation result, which confirms that cluster formation is minimized. The 12N+CNRs+ 5Y doped system shows more stability compared to 12N+CNRs+5Mn. The binding energy is stronger than the cohesive energy, and the distance between the Y-Y in the simulation result is also longer than the bulk Y-Y bond length as reported in [48]. The TM-doped systems are shown in Figure 3 (a) and (b). The dynamic stability of metal-doped 12N+CNRs is further confirmed by the phonon spectrum given in Figure 4. The absence of imaginary phonon modes in the negative frequency axis confirms that Y-decorated 12N+CNRs structures are dynamically stable. The Mn-functionalised 12N+CNRs exhibit a small imaginary phonon frequency (~−17 $cm^{-1}$), suggesting that any dynamical instability, if present, is weak and could be associated with numerical effects. There is a presence of a spectrum at lower frequencies for both Mn- and Y-doped 12N+CNRs, indicating heavy-metal vibrations, which are a sign of successful metal binding without disturbing the CNRs frameworks[49, 50].

Table 1. The magnitude of binding energy compared with the cohesive energy from the literature and DFT simulated result.

| System + TM | Simulated TM-TM distance (Å) | Literature bulk TM-TM distance (Å) | Binding energy (eV/Atom) | Cohesive Energy (eV/atom) |
|---|---|---|---|---|
| 12N+CNRs + 5Mn | 7.13 | 2.25 [47] | -2.57 | 2.92 [51, 52] |
| 12N+CNRs + 5Y | 6.97 | 3.4 [48] | -4.85 | 4.37 [52] |

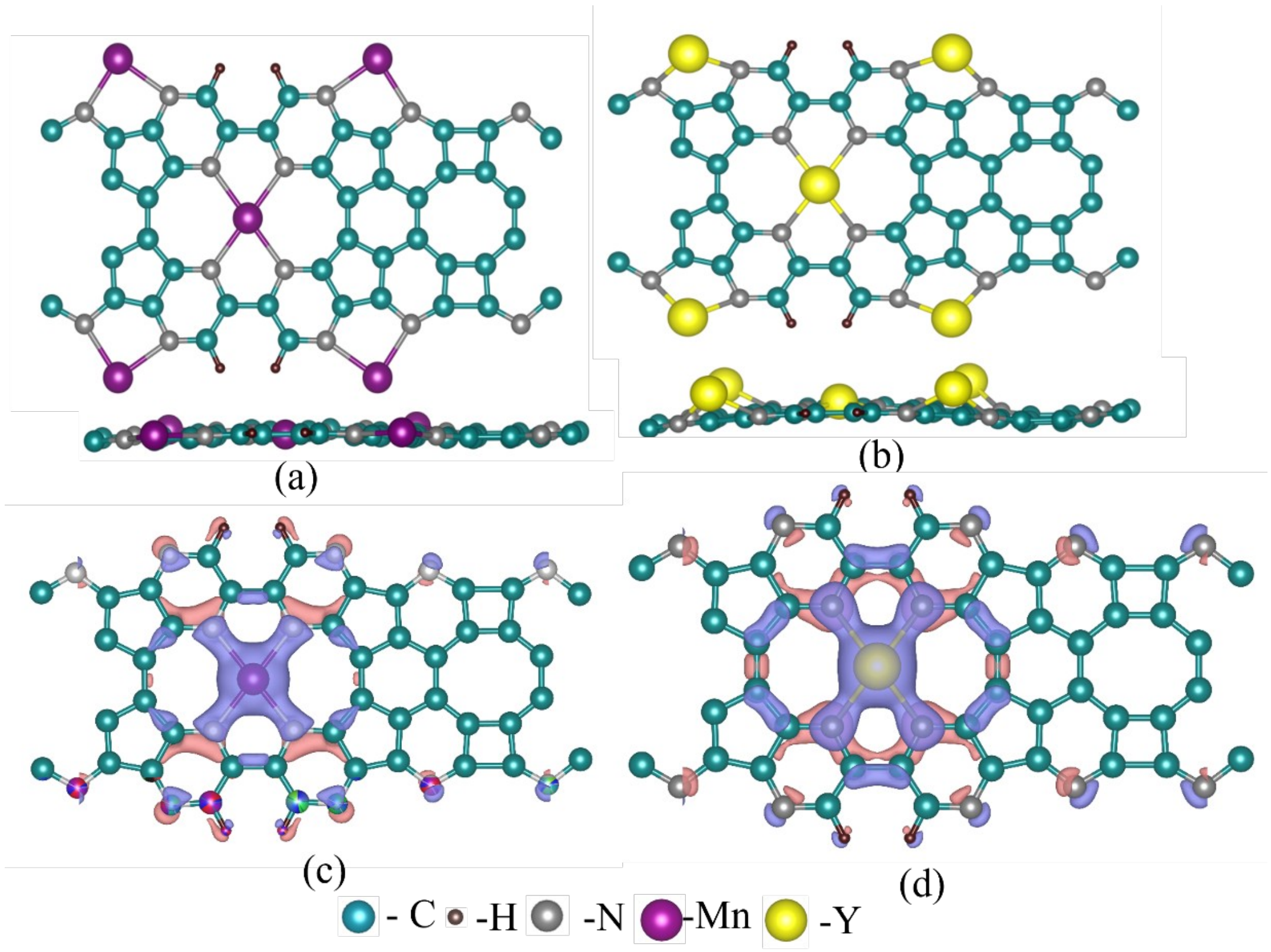

Figure 3. Final structure of transition metal-doped 12N+CNRs (a) 12N+CNRs+5Mn system, (b) 12N+CNRs+5Y system. (c) and (d) Charge density difference of 12N+CNRs+5Mn and 12N+CNRs+5Y systems, respectively. The purple region indicates charge accumulation; the yellow region indicates charge depletion, respectively with iso value 0.02 e/Å$^3$.

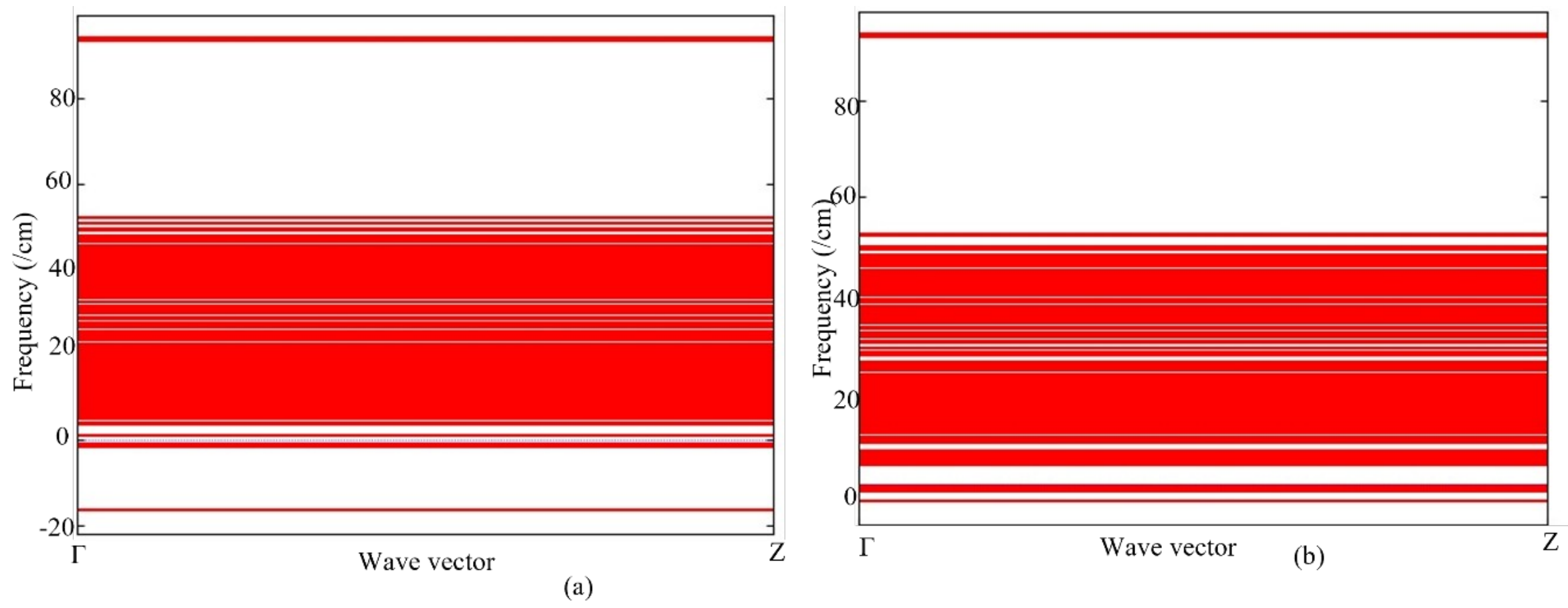

Figure 4. Phonon spectrum of (a) 12N+CNRs+5Mn (b) 12N+CNRs+5Y

The electronic properties of 12N+CNRs+5TM were analysed through PDOS as shown in Figure 2 (c) and (d). The 12N-CNR+5TM exhibits a transition from semiconducting to metallic upon transition metal decoration. The presence of finite electronic states at fermi level in two systems indicates strong orbital hybridization between host atoms and TM dopants. Both TM dopants show asymmetric PDOS between spin-up and spin-down. The partially filled d orbital is the primary reason where spin up orbital contains more electrons compared to spin down electrons [53], introducing a magnetic signature in the system. The induced magnetic moments are 1.02 $\mu_B$ and 5.77 $\mu_B$ for 12N+CNRs+Y and 12N+CNRs+Mn systems, close to the number of unpaired d electrons in the outermost shell. When 5 transition metals are introduced, the induced magnetic moments are 3.36 $\mu_B$ and 23.65 $\mu_B$ for 12N+CNRs+5Y and 12N+CNRs+5Mn systems due to magnetic interactions between the attached TM atoms. In both cases, the valence band region and conduction band region, the orbital overlapping is strongly contributed by C(p), N(p), Mn(d), and Y(d), with minimal contribution from passivated $H_2$ atoms.

The quantitative analysis of charge transfer using Bader charge calculation was conducted on transition metal doped 12N+CNRs system [23]. The results indicate that Mn and Y atoms can donate 1.24 and 2.12 electrons respectively to 12N+CNRs, exhibiting maximum charge transfer from the dopants to the substrate. This is an indication of dopants inducing a partial positive charge on TM dopants to polarize the $H_2$ molecules. The charge transfers are taking place due to the difference in electronegativity of different elements. The TM elements act as charge donors while the host material acts as a charge acceptor due to its high electronegativity. These charge donations of the two different systems are shown through charge density differences in Figure 3 (c) and (d). The charge accumulation and depletion are represented by the purple and yellow regions, respectively.

**3.3 Hydrogenation of 12N+CNRs+5TM**

After confirming the structural stability of the newly doped system, this section thoroughly discusses hydrogenation on the 12N+CNRs+5TM system. The hydrogenation first started with 5 $H_2$ molecules, 1 each on each TM metal dopant, and adsorption energies are calculated using Eq. ( 2). To consider the next loading of $H_2$ molecules, the adsorption energy is compared with US DOE standards[45]. If the $E_{ads}$ falls within the range of -0.20 to -0.6 eV/$H_2$, and structures are not deformed, the next loading of $H_2$ is considered. A total of 40 $H_2$ molecules were loaded to the TM-doped 12N+CNRs system with adsorption energies ranging from -0.20 to -0.60 eV/$H_2$. The top views and side views of the optimized hydrogenated structure of the TM-doped 12N+CNRs system are shown in Figure 5.

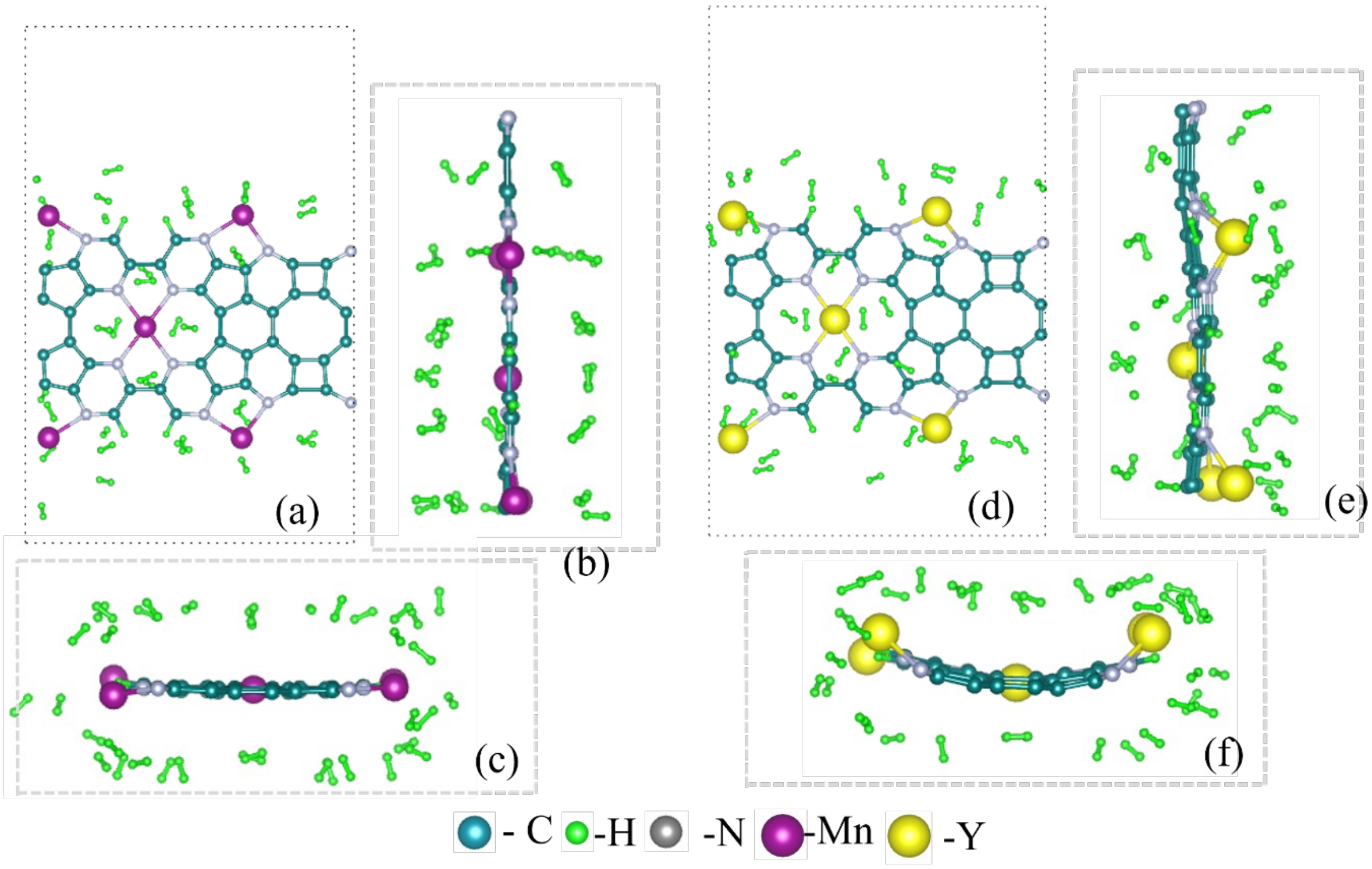

Figure 5. Optimized hydrogenated structure of (a) top view, (b) and (c) side views of 12N+CNRs+5Mn. (d) Top view, (e) and (f) side views of 12N+CNRs+5Y. The H atom legend is represented in green, indicating adsorbed $H_2$ molecules.

Initially, the $H_2$ molecules were placed 2 Å away from the surface of the 12N+CNRs+5TM systems with an H–H bond length fixed at 0.74 Å. After structural optimization, a slight elongation of the H–H bond length was observed, reaching up to 0.85 Å, which is consistent with the value reported in [23]. The interaction of $H_2$ with TM is due to Kubas-type interactions, where there is charge donation and back-donation between the H 1s orbital and TM's d orbital. There is a net gain in charge by $H_{2,}$ which causes elongation in the H-H bond length, but it is still in molecular form. The optimized structures showed that the $H_2$ molecules were adsorbed within a maximum distance of 3.5 Å from the doped 12N+CNRs surface. The average adsorption energies were calculated to be −0.40 eV/$H_2$ and −0.27 eV/$H_2$ for the 12N+CNRs+5Mn and 12N+CNRs+5Y systems, respectively. For each system, 40 $H_2$ molecules were initially loaded. After structural relaxation, 37 $H_2$ molecules for 12N+CNRs+5Mn and 38 $H_2$ molecules for 12N+CNRs+5Y remained within 3.5 Å of the 12N+CNRs surface and were considered adsorbed, while the remaining 3 and 2 molecules, respectively, were located beyond this distance and excluded from the storage-capacity calculations. The corresponding gravimetric $H_2$ storage capacities are 7.48 and 6.55 wt%, respectively. Both systems achieved storage capacities beyond the US DOE target, indicating their potential as promising $H_2$ storage materials. The detailed adsorption energies and corresponding $H_2$ storage capacities for different $H_2$ loading stages are presented in Figure 2S and Figure 3S in the supplementary document. The results of similar studies on nanomaterials for $H_2$ storage are reviewed and presented in Table 2.

Table 2. Adsorption energies and gravimetric weight capacity of different $H_2$ storage systems compared.

| 2D storage materials | Adsorption Energy (eV) | Gravimetric weight (wt%) |
|---|---|---|
| 12N+CNRs+5Mn, 12N+CNRs+5Y(this work) [a] | -0.40, -0.27 | 7.48, 6.55 |
| 12N+CNRs+(Li, Na, K, Mg, Ca) [43][a] | Ranges from -0.597 to -0.11 | Ranges from 7.08 to 3.74 |
| 12N+CNRs+(Sc and Ti)[30][c] | -0.28 and -0.29 | 6.10 and 5.49 |
| Li-g-$C_6N_7$[54][b] | -0.145 to -0.196 | 11.94 |
| Mg-g-$C_6N_7$[55][a] | -0.178 | 10.00 |
| Li-$Ti_2CF_2$[56][a] | -0.17 | 3.80 |
| Li-$BC_2P$ [57] | -0.20 to -0.28 | 6.68 |
| Li-C18[58][b] | -0.279 | 9.92 |
| Li-s-$C_3N_6$[26][b] | -0.22 | 7.84 |
| Sc, Ti, Fe, Cu doped g-$C_3$N4[59][a] | -0.33 to -0.54 | 1.43 |
| Li-$GeC_5$[60][b] | -0.22 | 7.62 |
| $C_3N_3$[61][c] | -0.25 | 16.13 |
| Li-$B_7N_5$[62][b] | -0.23 | 8.77 |
| Ca, Mg, K doped $C_3N_2$[23][a] | -0.24, -0.24, -0.19 | 7.53, 9.47, 8.21 |
| Li-$Be_2C$[63][a] | -0.193 | 10.40 |
| Mg-g-$C_2N$ [64][a] | -0.12 | 6.79 |
| 4% strain 3NHGY [65][b] | -0.191 to -0.338 | 6.42 |
| Ti-AzaCOF [66][b] | -0.43 | 9.34 |
| $B_2CO$ [50][a] | -0.614 | 5.94 |
| Li-$B_3O_3$, Na-$B_3O_3$[67][a] | -0.32 | 16.09, 10.07 |
| Li, Na, K doped $B_4CN_3$[68][a] | -0.07, -0.15, -0.11 | 7.13, 6.24, 6.83 |
| Be, Mg, Ca doped $B_4CN_3$[68][a] | -0.54, -0.25, -0.25 | 1.85, 4.70, 5.51 |
| Sc, Ti, V doped $B_4CN_3$[68][a] | -0.33, -0.46, -0.56 | 5.33, 5.22, 3.89 |
| $BeN_4$[69][a] | -0.157 | 7.19 |
| 10% compressed strain $BeB_4$ [70] | -0.201 | 17.72 |

[a] GGA-PBE/DFT-D3
[b] GGA-PBE/DFT-D2
[c] VDW-DF/DZP (double zeta polarised)

### 3.4 Thermodynamic Analysis.

#### 3.4.1 Relative energy and desorption temperature.

The stability of the hydrogenated 12N+CNRs+5TM system under practical operating pressure and temperature was evaluated using the relative energy ($E_r$) expression given in Eq. (9) [71].

$$E_r = E_\tau - E_\tau' - n\left(E_{H_2} - \mu(T,P)\right) \quad (9)$$

Where $E_\tau$ and $E_\tau'$ terms represent the calculated total energies of the hydrogenated and bare 12N+CNRs+5TM system, respectively, and $\mu(T,P)$ represents the temperature- and pressure-dependent chemical potential of $H_2$. The relative energy analysis was performed under fuel cell operating conditions, considering a temperature of 298.15 K and a $H_2$ pressure of 1 bar. The variation of $E_r$ with respect to pressure and temperature for the maximum $H_2$ loading configuration is presented in Figure 6. Figure 6 (a) illustrates the pressure-dependent relative energy behaviour at a constant temperature of 298.15 K. At very low pressures, the calculated $E_r$ values are positive, indicating that $H_2$ adsorption is thermodynamically unfavourable due to the low chemical potential of $H_2$. As the pressure gradually increases, $E_r$ gradually decreases and becomes negative, suggesting that $H_2$ adsorption becomes energetically favourable under higher pressure conditions. This indicates that the 12N+CNRs+5TM systems can effectively stabilize the adsorbed $H_2$ molecules at elevated pressures. Figure 6 (b) presents the temperature dependence of $E_r$ at a fixed pressure. The relative energy remains negative at lower temperatures, confirming the favourable adsorption of $H_2$ molecules. However, with increasing temperature, $E_r$ shifts towards positive values due to the increase in $H_2$ chemical potential, which reduces the thermodynamic driving force for adsorption. Therefore, the relative energy analysis confirms that $H_2$ storage in the 12N+CNRs+5TM systems is thermodynamically more favourable under high-pressure and low-temperature conditions, which are consistent with the operating principles of $H_2$ storage materials [43].

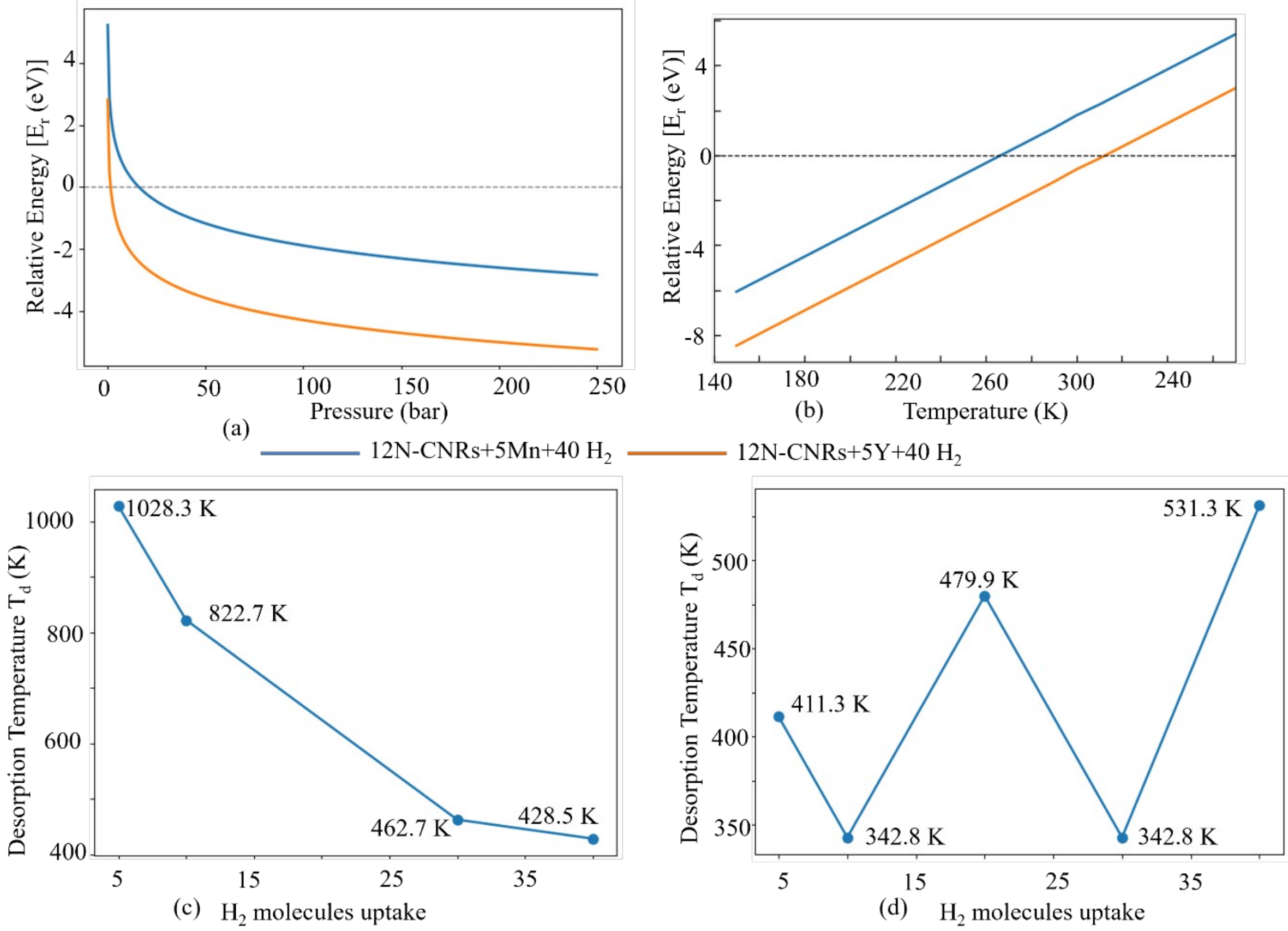

Figure 6. Relative energy $E_r$ of the maximum hydrogenated system as a function of (a) pressure at fixed temperature (298.15 K), (b) temperature at fixed pressure (10 bar), (c) and (d) desorption temperature for varying $H_2$molecule uptake for 12N+CNRs+5Mn and 12N+CNRs+5Y, respectively.

To investigate the dynamics of $H_2$ adsorption on the 12N+CNRs+5Mn and 12N+CNRs+5Y systems, the $H_2$ desorption temperature ($T_d$) was analysed and is presented in Figure 6 (c) and (d), respectively, using Eq. (5). The adsorption energy decreases with increasing $H_2$ uptake, resulting in a gradual reduction in the desorption temperature for the 12N+CNRs+5Mn system. However, a non-monotonic variation in the desorption temperature is observed for the 12N+CNRs+5Y system, which can be attributed to the heterogeneous adsorption energies of $H_2$ molecules at different adsorption sites. Although some fluctuations are observed, the desorption temperature remains within a moderate range, indicating reversible $H_2$ adsorption under practical operating conditions.

### 3.4.2 Reversible storage of $H_2$ under varying temperature and pressure.

Comparing with previous literature of [30, 43] of similar CNRs studies, the current study shows adequate adsorption energy and more storage capacity at 0.0 K. To consider the practical application of storage materials, with the help of the grand canonical partition function given in Eq. (6), (7), and (8), the $H_2$intake was calculated under varying temperature and pressure.

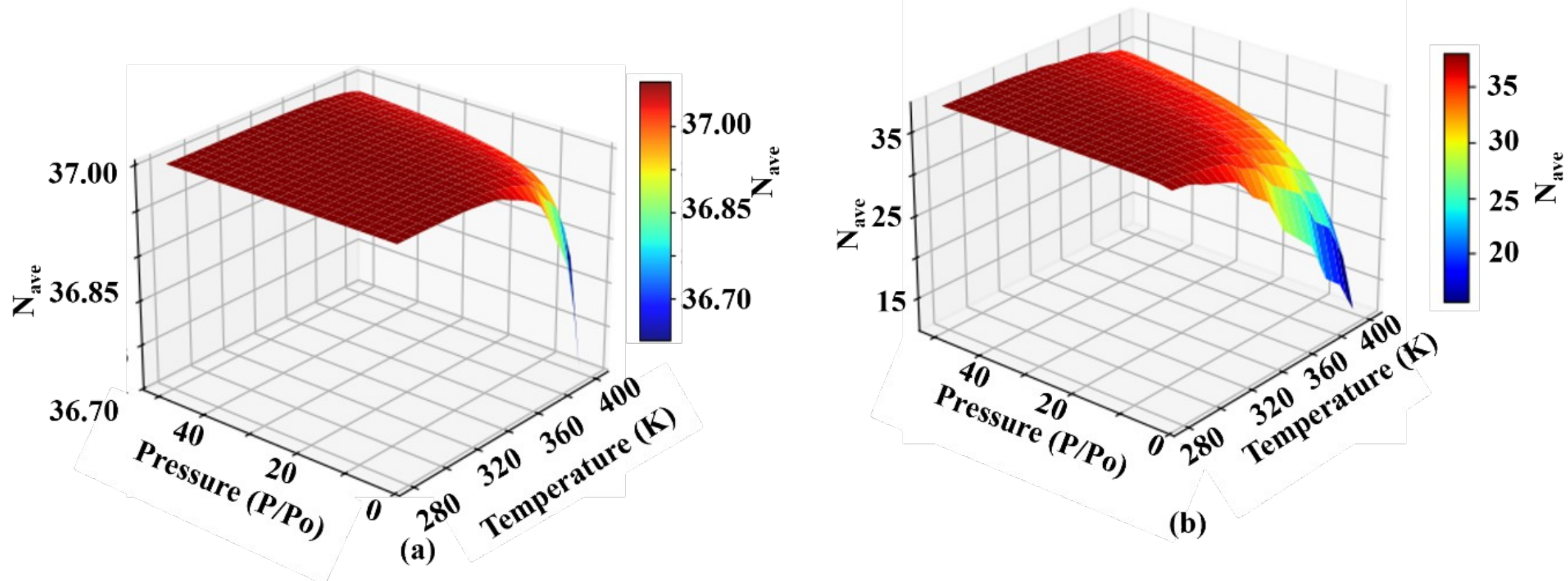

Figure 7. Average number of $H_2$ uptake on (a) 12N+CNRs+5Mn and (b) 12N+CNRs+5Y at varying temperature and pressure.

The thermodynamic $H_2$ storage performance of the 12N+CNRs+5TM systems was further evaluated under practical operating conditions of 298.15 K and 30 atm, representing realistic temperature and pressure environments for $H_2$ storage applications. At 0 K, the maximum $H_2$ storage capacities were calculated to be 7.48 wt% for the 12N+CNRs+5Mn system and 6.55 wt% for the 12N+CNRs+5Y system. After incorporating the temperature and pressure effects, the $H_2$ storage capacity of the 5Mn-doped system remained unchanged at 7.48 wt%, while the capacity of the 5Y-doped system decreased slightly to 6.04 wt%. This reduction is attributed to the influence of thermal energy, where increasing temperature promotes $H_2$ desorption by weakening the interaction between 12N+CNRs and TMs, whereas increasing pressure enhances $H_2$ adsorption by increasing the chemical potential of gaseous $H_2$ and driving more molecules toward the 12N+CNRs surface. Therefore, the thermodynamic evaluation under practical conditions confirms the potential of both doped systems for reversible $H_2$ storage applications.

## Conclusion

In this study, transition-metal-functionalised 12N+CNRs were investigated for $H_2$ storage using DFT calculations. Incorporation of Mn and Y provided stable anchoring sites and enhanced $H_2$ adsorption through strong electronic interactions, as confirmed by PDOS and Bader charge analyses. The calculated metal binding energies of −2.57 and −4.85 eV per dopant for Mn and Y, respectively, exceeded their cohesive energies, indicating that metal clustering is not favourable. The systems exhibited favourable average $H_2$ adsorption energies of −0.40 eV/$H_2$ for Mn and −0.25 eV/$H_2$ for Y, within the desirable range for reversible $H_2$ storage. The maximum gravimetric storage capacities reached 7.48 wt% for Mn and 6.55 wt% for Y at 0 K. Under practical conditions of 30 atm and 298.15 K, the Mn- and Y-doped systems retained capacities of 7.48 and 6.04 wt%, respectively. Thermodynamic analysis further indicated favourable $H_2$ adsorption at lower temperatures and higher pressures, with desorption promoted by increasing temperature and decreasing pressure. Overall, both systems demonstrate promising $H_2$ storage characteristics, with Mn-functionalised 12N+CNRs showing superior practical storage performance. These findings highlight the potential of TM-functionalised 12N+CNRs for $H_2$ storage, although further experimental validation is required.

Acknowledgement

The authors sincerely acknowledge the University of New England for providing the PhD scholarship that supported this research. The authors also acknowledge BARC's Supercomputing facility and its staffs for the support.